\documentclass[preprint, 3p, authoryear, 12pt]{elsarticle}
\usepackage{amsmath}
\usepackage{amsfonts}
\usepackage{amssymb}
\usepackage{subfigure}
\usepackage{color}
\usepackage[usenames,dvipsnames,svgnames,table]{xcolor}

\usepackage{graphicx,bm}
\makeatother

\begin{document}

\begin{frontmatter}

\title{Effect of size on necking of dynamically loaded notched bars}

\author{A. Needleman}

\address{Department of Materials Science \& Engineering,
 Texas A\&M University,  College Station, TX 77843 USA}

\begin{abstract}
The influence of material inertia on neck development in a notched round bar
is analyzed numerically. Dynamic axisymmetric calculations are
carried out for isotropically hardening elastic-viscoplastic solids so
that both material strain rate sensitivity and material inertia are
accounted for. The
focus is on the effect of bar size on whether the 
notch triggers necking or necking initiates away from the notch.  The
governing equations 
are presented in non-dimensional form and two key non-dimensional groups
that involve both material and loading parameters are identified. For
both non-dimensional groups, with all parameters fixed except for bar
size, it is found that for sufficiently small bars, the notch triggers
necking, whereas for sufficiently large bars necking ultimately occurs
away from the notch. With material properties fixed, for one
non-dimensional group the transition to 
necking away from the notch corresponds to increasing imposed velocity
whereas for the other non-dimensional group this transition takes place for
decreasing imposed strain rate. Both these transitions correspond to
increasing bar size.  The results indicate that this
transition is governed by material inertia but the bar size at which
it occurs depends on the material properties, particularly strain
hardening and strain rate hardening.
\end{abstract}


\end{frontmatter}

\section{Introduction}

There is an extensive, more than 100 year old,
literature on the mechanics of 
necking in the uniaxial tensile test. The classical criterion of
 \cite{C85} holds 
for necking of a tensile bar in the limiting case of a infinitely long, thin 
bar and states that necking initiates at the maximum
load. For any finite aspect ratio, there is a delay between the 
maximum load point and the onset of necking that increases as the bar
becomes more stubby, \cite{N72,HM74,HN77}. The
literature on the analyses of necking in tensile bars
includes one dimensional analyses as well as full three dimensional finite element
solutions, involving both quasi-static and dynamic formulations, and analyses
that account for effects of various mechanical
properties, such as thermal softening, porosity induced softening, bar
geometry, etc. Reviews of tensile bar necking
analyses are provided by \cite{JWH79,MMJ14}. 

For rate independent plasticity and quasi-static deformations, the
onset of necking in a uniform circular cylindrical tensile bar is
associated with a 
bifurcation from a state of homogeneous uniaxial tension
\cite{CAD71,N72,HM74}. The bifurcation mode is 
associated with a sinusoidal variation in the radial dimension of the
bar with the longest possible wavelength consistent with the bar
geometry and the boundary conditions at the bar ends. A geometrical
imperfection leads to the fairly abrupt development 
of this mode at an overall strain somewhat less (depending on the
imperfection) then the bifurcation strain. Once the neck develops, the
classic approximate analysis of \cite{B52}, and subsequent
full numerical solutions, e.g. \cite{C71,N72,Im75,norris78,TN84,TN85}, show
that the neck curvature induces stress triaxiality that plays a key
role in the ductile failure process.

For a viscoplastic solid under
quasi-static loading conditions, the 
onset of necking is no longer associated 
with a bifurcation. However, the onset of necking can be analyzed as
the growth of an initial inhomogeneity, \cite{HN77}. As for a rate
independent plastic solid, a notch serves as an imperfection that
triggers necking and sets the neck location. Material rate
sensitivity leads to a delay in the onset of necking,  \cite{HN77}.

In addition, for both rate
independent and rate dependent plastic solids characterized by a
classic plastic constitutive relation, there is no material 
length scale in a quasi-static analysis. Hence, the evolution of the 
neck with strain (at the same imposed strain rate for viscoplastic solids) is
independent of specimen size. 

The necking behavior under
dynamic loading conditions, e.g.  
\cite{AN91,KN93,FN94,GF02,MN03,rusinek05,Os13,VR15,ROR15}, can be quite
different than under quasi-static conditions. Material inertia tends to slow neck
development, \cite{AN91,Xue08}; multiple necking can occur,
e.g. \cite{KN93,FN94,GF02}; there are size effects,
e.g. \cite{rusinek05,KN93}, and neck 
development can ignore the presence of notches, \cite{ROR15}. Experiments and
modeling carried out in \cite{ROR15} showed that under dynamic loading
conditions, the onset of necking in notched tensile bars could occur
away from the notch location.  

Since material inertia implicitly introduces a length scale,
different size specimens deformed at the same strain rate
may respond differently. As a consequence, there can be a
dependence of the failure strain on specimen size, \cite{KN93}.
\cite{KN93}  carried out finite deformation
dynamics analyses 
aimed at modeling the effect of specimen size at a
fixed imposed strain rate, on ductile failure in
geometrically self-similar tensile bars having various
sizes. The material was 
modeled as a viscoplastic progressively cavitating solid. It was found
that the variation of the necking strain with specimen size was not
monotonic; the response of sufficiently small specimens was
essentially quasi-static and size independent, the
failure strain then increased with specimen size before eventually
decreasing for sufficiently large specimens.

In this study, a combination of the
issues addressed in \cite{KN93} and \cite{ROR15} is considered. In
particular, a main 
focus in this paper is to continue exploring the issue raised by
\cite{ROR15} concerning the circumstances, for
dynamic loading conditions, under which necking ignores the presence
of a notch. 

Calculations are carried out for geometrically
similar, dynamically loaded notched
circular cylindrical tensile bars of various sizes. Attention is restricted to
axisymmetric deformations. The bar material is characterized as an
isotropically hardening viscoplastic Mises solid. A non-dimensional
form of the governing equations is presented and two key
non-dimensional ratios are identified: one relates the bar length
to a characteristic length that depends on material properties and the
imposed velocity, while the other relates the imposed strain
rate (the imposed velocity divided by the bar length) to a material
characteristic strain rate.   Both of these non-dimensional ratios involve
the bar length. The focus of the results is on the transition from
necking at the notch cross section to necking away from the notch
cross section as the specimen size is varied. 

\section{Problem Formulation}

As in \cite{KN93}, the calculations are based on a convected
coordinate Lagrangian formulation of the field
equations.  The independent variables are taken to be the particle
positions in the initial stress free configuration of the axisymmetric
tensile bar and time. In the current configuration the material point
initially at ${\bf X}$ is at ${\bf x}$.  The displacement vector ${\bf
  u}$ and the deformation gradient ${\bf F}$ are defined by 
\begin{equation}
{\bf u}= {\bf x} - {\bf X} \quad , \quad {\bf F}= \frac{\partial {\bf u}}{\partial {\bf X}}
\end{equation}
The principle of virtual work accounting for material inertia is written as 
\begin{equation}
\int_V  {\bf S}:\delta {\bf F} dV=\int_B ( {\bf S} \cdot  {\bf n}) \cdot \delta  {\bf u} dB
-\int_V \rho  \frac{\partial^2 {\bf u}}{\partial T^2} \cdot \delta
{\bf u} dV
\label{pvw}
\end{equation}
Here, $T$ is time, ${\bf S}$ is the (unsymmetric) nominal stress tensor, $ {\bf S}=
({\rm det} {\bf F}) {\bf F}^{-1} \cdot \boldsymbol{\Sigma}$ with
$\boldsymbol{\Sigma}$ the Cauchy stress, $\rho$ is the mass density,
and $V$ and $B$ are, respectively,  the volume and the surface of the
body in the undeformed reference configuration.  

Attention is confined to axisymmetric deformations and, for
notational simplicity, we use $r$
and $z$ to denote the convected Lagrangian coordinates. 
The initial length of the bar is $2L_0$ and the initial radius, which
varies along the bar is denoted by $R_0(z)$. The bar occupies the
region $-L_0 \le z \le L_0$, $0 \le r \le R_0(z)$.

An axial velocity $V(t)$ is imposed on $z=L_0$ together with
symmetry about $z=0$.  This means that the loading is actually applied
at $z=-L_0$ as well. The reason for imposing symmetry about $z=0$ is that
without this symmetry and with shear free conditions on the loading
ends, the preferred quasi-static necking mode would be the long
wavelength mode with the neck forming at one of the ends. Thus, under
quasi-static loading conditions the deformation and stress
concentrations associated with the centrally placed notch would be
competing with those associated with the preferred necking mode. On
the other hand, if shear constraints were imposed at the ends, then
the onset of necking would be affected by
both the notch and the deformation gradient imposed by the
constraints, complicating the interpretation of the notch effect. With
symmetry about $z=0$ imposed, the preferred quasi-static necking mode
is driven by the the presence of the notch and necking occurs
at the notch cross section. The occurrence of 
necking away from the notch cross section, when it occurs, is a dynamic
effect.  

The boundary conditions on the region analyzed are $u_z(r,L_0,T)=V(T)$ where
\begin{equation}
V(T)=\begin{cases}
 V_1 \, T/T_r, &\mbox{for } T \le T_r \\
 V_1 ,&\mbox{for } T > T_r
\end{cases}
\end{equation}
Here, $V_1$ is the magnitude of the imposed velocity and $T_r$ is the rise time.

The other displacement boundary conditions imposed are $u_z(r,0,T)=0$
and $u_r(0,z,T)=0$. All other boundary 
conditions correspond to zero imposed tractions.  

The material is characterized as an elastic-viscoplastic Mises solid.
The total rate of deformation, ${\bf D}$, is written as the sum of an
elastic (actually hypoelastic) part, ${\bf D}^e$, and a viscoplastic part, ${\bf D}^p$, with 
\begin{equation}
{\bf D}^e=\frac{1+\nu}{E} {\tilde {\boldsymbol{T}}}-
\frac{\nu}{E} {\rm tr}(\tilde{ {\boldsymbol{T}}}){\bf I} 
\label{con1}
\end{equation}
where $E$ is Young's modulus, $\nu$ is Poisson's ratio,
$\boldsymbol{T}=({\rm det} {\bf 
  F}) \boldsymbol{\Sigma}$, $(\tilde{\
})$ denotes the Jaumann rate based on $T$, ${\rm tr}( \ )$ denotes the trace and
${\bf I}$ is the identity tensor.

The viscoplastic flow rule is
\begin{equation}
{\bf D}^p=\frac{3 \dot {\Lambda}^p}{2 \Sigma_e}
\boldsymbol{T}^\prime
\label{con2}
\end{equation}
where ${\bf I}$ is the identity tensor, $\dot {\Lambda}^p$ is the
effective plastic strain rate, and the Kirchhoff stress deviator
$\boldsymbol{T}^\prime$ and effective stress $ \Sigma_e$ are given by 
\begin{equation}
\boldsymbol{T}^\prime=\boldsymbol{T}- \Sigma_h {\bf I} \quad , \quad  \Sigma_e=
\sqrt{\frac{3}{2}\boldsymbol{T}^\prime : \boldsymbol{T}^\prime} \quad ,
\quad \Sigma_h=\frac{1}{3} {\rm tr}(\boldsymbol{T}) {\bf I} 
\end{equation}

The material response, the specimen geometry and the boundary value
problem are characterized by a collection of non-dimensional
quantities. To put the equations in non-dimensional form, we normalize
all stress quantities by a reference stress $\sigma_0$, all length
quantities by a reference length $L_c$ and all time quantities by a
reference time $t_c$. 

The principle of virtual work, Eq.~(\ref{pvw}), can be written as
\begin{equation}
\int_v  {\bf s}:\delta {\bf F} dv=\int_b  ({\bf s} \cdot {\bf n}) \cdot \delta  {\bf w} db
-\int_V  \ddot{\bf w} \cdot \delta {\bf w} dv
\label{pvw2}
\end{equation}
provided
\begin{equation}
t_c={L_c}\sqrt{\frac{\rho}{\sigma_0}}
\label{rel1}
\end{equation}
In Eq.~(\ref{pvw2}), ${\bf s} =\sigma_0 {\bf S}$, ${\bf u} = L_c {\bf w}$, $\ t=t_c T$,
$dV=L_c^3 dv$, $dB=L_c^2 db$ and $(\dot \ )$ denotes $\partial (
\ )/\partial t$.

As in \cite{KN93}, we take
\begin{equation}
L_c=L_0 \frac{c_0}{V_1} \ , \ c_0=\sqrt{\frac{E}{\rho}}
\label{rel2}
\end{equation}
Hence, from Eq.~(\ref{rel1})
\begin{equation}
t_c=\frac{L_0}{V_1}\sqrt{\frac{E}{\sigma_0}}
\label{rel3}
\end{equation}

In non-dimensional form, the rate constitutive relation,
Eqs.~(\ref{con1}) and (\ref{con2}) become
\begin{equation}
{\bf d}^e=\epsilon_0\left [ (1+\nu) {\hat {\boldsymbol{\tau}}}-
\nu {\rm tr}(\hat{ {\boldsymbol{\tau}}}){\bf I} \right ]
\quad , \quad \epsilon_0=\frac{\sigma_0}{E}
\label{con3}
\end{equation}
where $(\hat{\ })$ denotes the Jaumann rate based on $t$, and
\begin{equation}
{\bf d}^p=\frac{3 \dot {\epsilon}^p}{2 \sigma_e}
\boldsymbol{\tau}^\prime
\label{con4}
\end{equation}
In Eq.~(\ref{con4}), $\dot {\epsilon}^p=t_c \dot {\Lambda}^p$.

The plastic response of the material is characterized by a power law
rate hardening of the form  
\begin{equation}
\dot{\epsilon}_p=t_c \dot{\epsilon}_0 \left( \frac{\sigma_e}{g} \right)^{1/m} 
\label{ep-def}
\end{equation}
Here, $\sigma_{e}=\Sigma_e/\sigma_0$, $\dot{\epsilon}_0$ is a
reference strain rate, $m$ is the rate  
sensitivity exponent.  

The flow strength function
$g$ in Eq.~(\ref{ep-def}) is taken to be a function of $\epsilon_p$
and have the form
\begin{equation}
g\left(\epsilon_p\right)=  \left [ 1+\frac{\epsilon_p}{\epsilon_0} \right ]^N
\label{hardening}
\end{equation}
Here, $\epsilon_p=\int \dot{\epsilon}_p dt$ and $N$ is the strain hardening
exponent. Since $g(0)=1$ in Eq.~(\ref{hardening}), $\sigma_0$ is
now identified with the flow strength at zero plastic strain. Note that up
to the identification in Eq.~(\ref{hardening}), $\sigma_0$ could be any
convenient quantity having the dimension of stress.

The constitutive response is characterized by the non-dimensional
material parameters. 
$\epsilon_0$, $\nu$, $m$ and $N$. For a given function
$R_0(z)$, the specimen geometry is characterized by the non-dimensional
ratio $R_0(0)/L_0$.

There are two key non-dimensional groups involving both loading and material
parameters. One is 
\begin{equation}
\frac{L_0}{L_c}=\frac{V_1}{c_0}
\label{eq:L}
\end{equation}
and the other is
\begin{equation}
\kappa=\frac{V_1/L_0}{\dot{\epsilon}_0}
\label{eq:kap}
\end{equation}
which is the ratio of the imposed strain rate to the 
material characteristic strain rate. The ratio $L_0/L_c$ provides a
measure of the effect of material inertia while $\kappa$ provides a measure of
the effect of loading rate, independent of material inertia. In the quasi-static limit
$L_0/L_c \rightarrow 0$ so that, of course, the effect of inertia on
the response vanishes. On the other hand, the role of $\kappa$
persists in the quasi-static limit.

There are other non-dimensional groups relating
material parameters and loading parameters, but they are not
independent of the ones already defined. For example, 
$t_c \dot{\epsilon}_0$, which can be regarded as a ratio of 
dynamic and constitutive time scales is, from Eqs.~(\ref{rel3}) and
(\ref{eq:kap}),  given by 
\begin{equation}
t_c \dot{\epsilon}_0=\frac{L_0
  \dot{\epsilon}_0}{V_1}\sqrt{\frac{E}{\sigma_0}}
=\frac{1}{\kappa \sqrt{\epsilon_0}} 
\label{eq:ratx}
\end{equation}

Eq.~(\ref{ep-def}) then can be rewritten as 
\begin{equation}
\dot{\epsilon}_p=\frac{1}{\kappa \sqrt{\epsilon_0}} \left(
  \frac{\sigma_e}{g} \right)^{1/m}  
\label{ep-def2}
\end{equation}
and used to express the plastic flow rule,
Eq.~(\ref{con4}) in non-dimensional form as
\begin{equation}
{\bf d}^p=\frac{3}{2} \frac{1}{\kappa \sqrt{\epsilon_0}} \left(
  \frac{\sigma_e}{g} \right)^{1/m}  \left(
  \frac{\boldsymbol{\tau}^\prime}{ \sigma_e} \right )
\label{con5}
\end{equation}
Eq.~(\ref{con5}) explicitly shows the constitutive dependence on the
non-dimensional parameter $\kappa$ which gives rise to a coupling
of the constitutive response to parameters involving the bar geometry
and the imposed loading. This coupling occurs because of
material rate sensitivity and does not occur for the corresponding
rate independent  plastic flow rule.

Another non-dimensional ratio is the ratio of the stress carried by the
loading wave to the reference stress. A one
dimensional linear elastic wave propagation analysis of a tensile
bar subject to a prescribed end velocity $V_1$ gives the stress
carried by the loading wave as $\rho c_0 V_1$, see e.g. 
\cite{Lee67}. However, this is not an independent ratio. 
The ratio of this loading wave stress to the reference stress $\sigma_0$ is
\begin{equation}
\frac{\rho c_0 V_1}{\sigma_0}=\frac{1}{\epsilon_0}
\frac{V_1}{c_0}=\frac{1}{\epsilon_0} \frac{L_0}{L_c}
\label{rel4}
\end{equation}
Hence, the stress carried by the loading wave (according to a one
dimensional linear elastic analysis)  is proportional to $L_0/L_c$ and is
inversely proportional to the non-dimensional material parameter
$\epsilon_0$. Hence, with material properties fixed, the stress
carried by the loading wave 
increases with increasing relative bar size, $L_0/L_c$ with the
quasi-static limit emerging as $L_0/L_c \rightarrow 0$. 

Thus, the boundary value problem is characterized by the non-dimensional
material parameters, $\epsilon_0$, $\nu$, $N$ and $m$; the
non-dimensional geometry parameter $R_0/L_0$; the non-dimensional rise
time $t_r$; and the 
non-dimensional parameters, $L_0/L_c$ and $\kappa$, that involve
geometry, loading and material parameters.  

To illustrate the scaling, let $\rho^*=A \rho$, then with
\begin{equation}
E^*=A^p E \ , \ \sigma_0^*=A^p \sigma_0
\label{eq:sc1}
\end{equation}
the linear elastic wave speed scales as $c_0^*=A^{(p-1)/2}c_0$.

Taking $V_1^*=A^{(p-1)/2}V_1$, $L_0^*=A^{(p-1)/2}L_0$, and
$\dot{\epsilon}_0^*=\dot{\epsilon}_0$, the non-dimensional ratios
$L_0/L_c$, $\epsilon_0$ and $\kappa$, as well as the characteristic
time $t_c$ are unchanged with this scaling. Thus, with 
this scaling the solution of the dynamic 
initial/boundary value problem (with fixed $R_0/L_0$) coincides for
both these sets of material and loading parameters.

\section{Numerical Method and Results}
\label{result}

The numerical method is basically the same as in \cite{AN91,KN93}.  
The discretization is based on linear displacement triangular elements
arranged in quadrilaterals of four ``crossed'' triangles and the time
integration of the discretized governing equations are 
integrated numerically by an explicit integration procedure,
\cite{Belytschko76}, with a lumped mass matrix.
The constitutive update is based on the rate 
tangent modulus method of \cite{Peirce84}. 

\begin{figure}[htb!]
\begin{center}
\resizebox*{35mm}{!}{\includegraphics{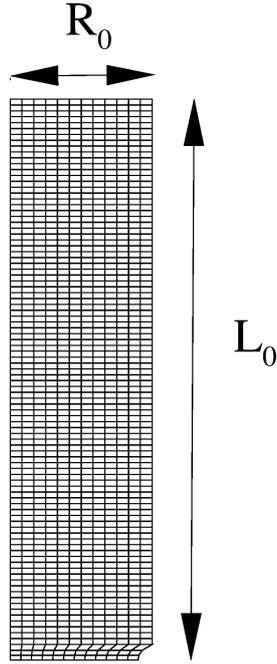}}
\end{center}
\caption{The $12 \times 96$ quadrilateral finite element mesh used in
  the computations. Each quadrilateral consists of four ``crossed''
  linear displacement triangles.}
\label{mesh}
\end{figure}

The fixed (non-dimensional) constitutive parameters are taken to be
$\epsilon_0=0.004$, $\nu=0.3$. The calculations are carried out for strain rate
hardening exponents $m=0.01$ and $m=0.05$. In most calculations the
strain hardening exponent is taken to be 
$N=0.01$, which is nearly ideally plastic, but a few calculations employ
$N=0.1$ to assess the effect of strain hardening. The value of the
(non-dimensional) rise time is taken to be $t_r=1.265\times 10^{-3}$ in
all calculations.

To give a perspective on what these non-dimensional parameter values
could correspond to, one possibility is $E=200$GPa, $\Sigma_0=800$MPa,
$c_0=5000$m/s, $V_1/L_0=1000$s$^{-1}$, $\dot{\epsilon}_0=1000$s$^{-1}$ and
$T_r=20 \times 10^{-6}$s. Also, with these values
$L_c=c_0/(V_1/L_0)=5$m (close to the value in \cite{KN93}). 

The bar aspect
ratio is fixed at $L_0/R_0=4$ and the bar has a notch as depicted in
Fig.~\ref{mesh}. The notch is a semi-circle of radius $R_0/10$
centered at $z=0$, $r=R_0$. However, as can be seen in Fig.~\ref{mesh}
there are only three nodal points on the notch surface so that the
circular shape is not faithfully represented. The notch mainly serves as an
imperfection to trigger necking at $z=0$. 

We define the ratio of current cross section area to initial cross
section area, $A_r(z)$, by
\begin{equation}
A_r(z)=\frac {\pi R_0^2(z)}{\pi [R_0(z)+u_r(R_0(z),z)]^2}
\label{area}
\end{equation}
A calculation is terminated when the maximum of $A_r(z)$ reaches
$2$ (an area reduction of $1/2$).  In the calculations here, this
occurs either for $A_r(0)$ or $A_r(L_0)$.  
If $A_r(0)$ reaches $2$ first then necking has occurred at the notch,
if $A_r(L_0)$ reaches $2$ first then necking has occurred away from
the notch and necking has ignored the presence of the notch. The value
$A_r=2$ is chosen arbitrarily. This value was chosen because neck
development has clearly occurred when $A_r=2$ yet the strains are not
so large that the finite element grid in Fig.~\ref{mesh} has become
significantly deformed. As will be seen subsequently, the value of
necking strain is not generally very sensitive to the precise cut-off value
chosen. 

At the termination of the calculation we define the ratio
\begin{equation}
R_A=\frac{A_r(0)}{A_r(L_0)}
\label{RA}
\end{equation}
so that $R_A>1$ implies necking at the notch, $R_A<1$ implies necking
away from the notch and $R_A=1$ corresponds to simultaneous necking at
the notch and away from the notch. Assuming a non-negative effective
Poisson's ratio, the possible range of values for $R_A$ is $0.5 \le
R_A \le 2$.

\subsection{Fixed $\kappa$, varying $L_0/L_c$}
\label{Lc}

In this section the
value of $\kappa$, Eq.~(\ref{eq:kap}), is fixed at
$\kappa=1$ and $L_0/L_c$, Eq.~(\ref{eq:L}), is varied.

\begin{figure}[htb!]
\begin{center}
\subfigure[]
{\resizebox{!}{60mm}{\includegraphics{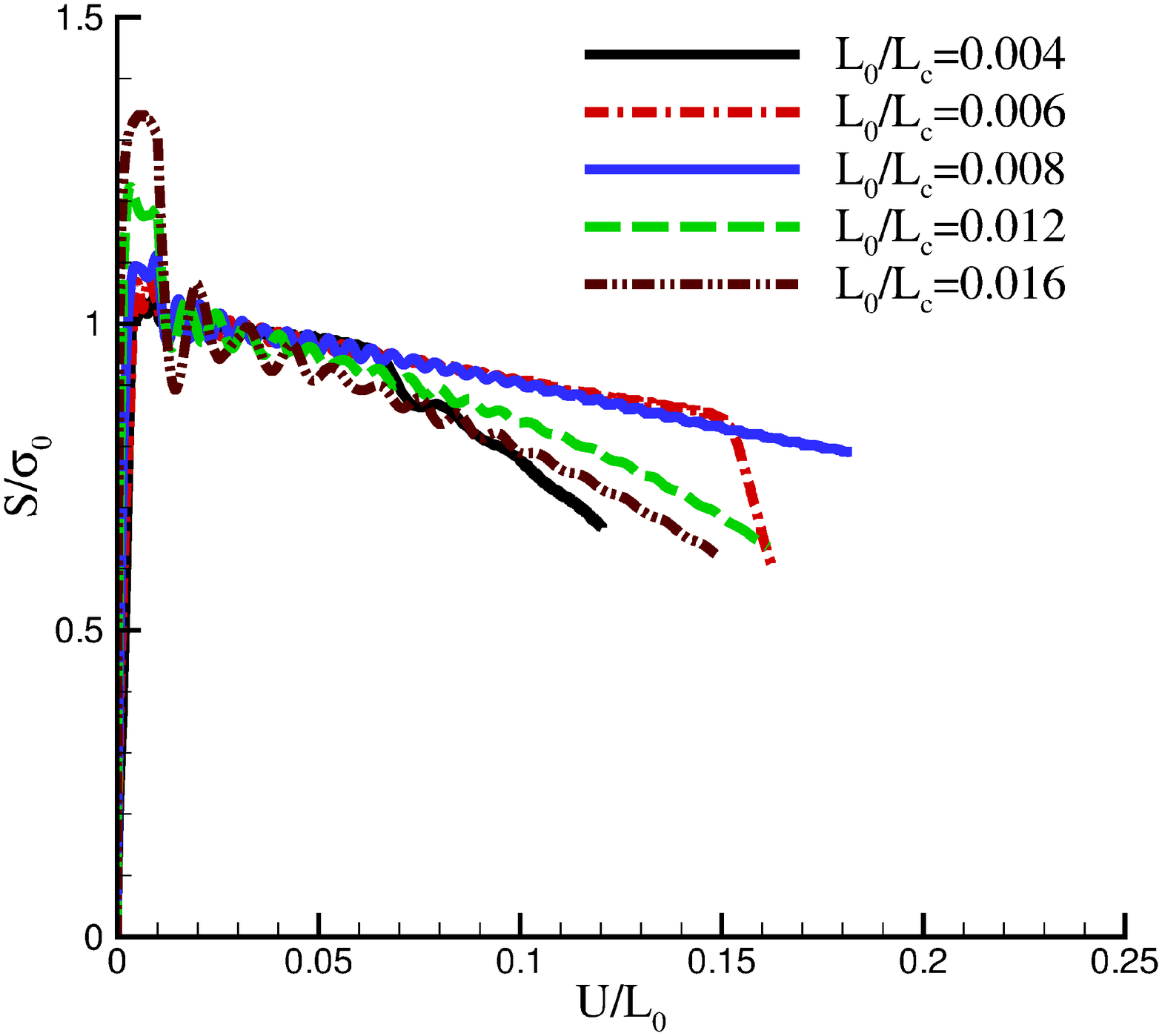}}}
\subfigure[]
{\resizebox{!}{60mm}{\includegraphics{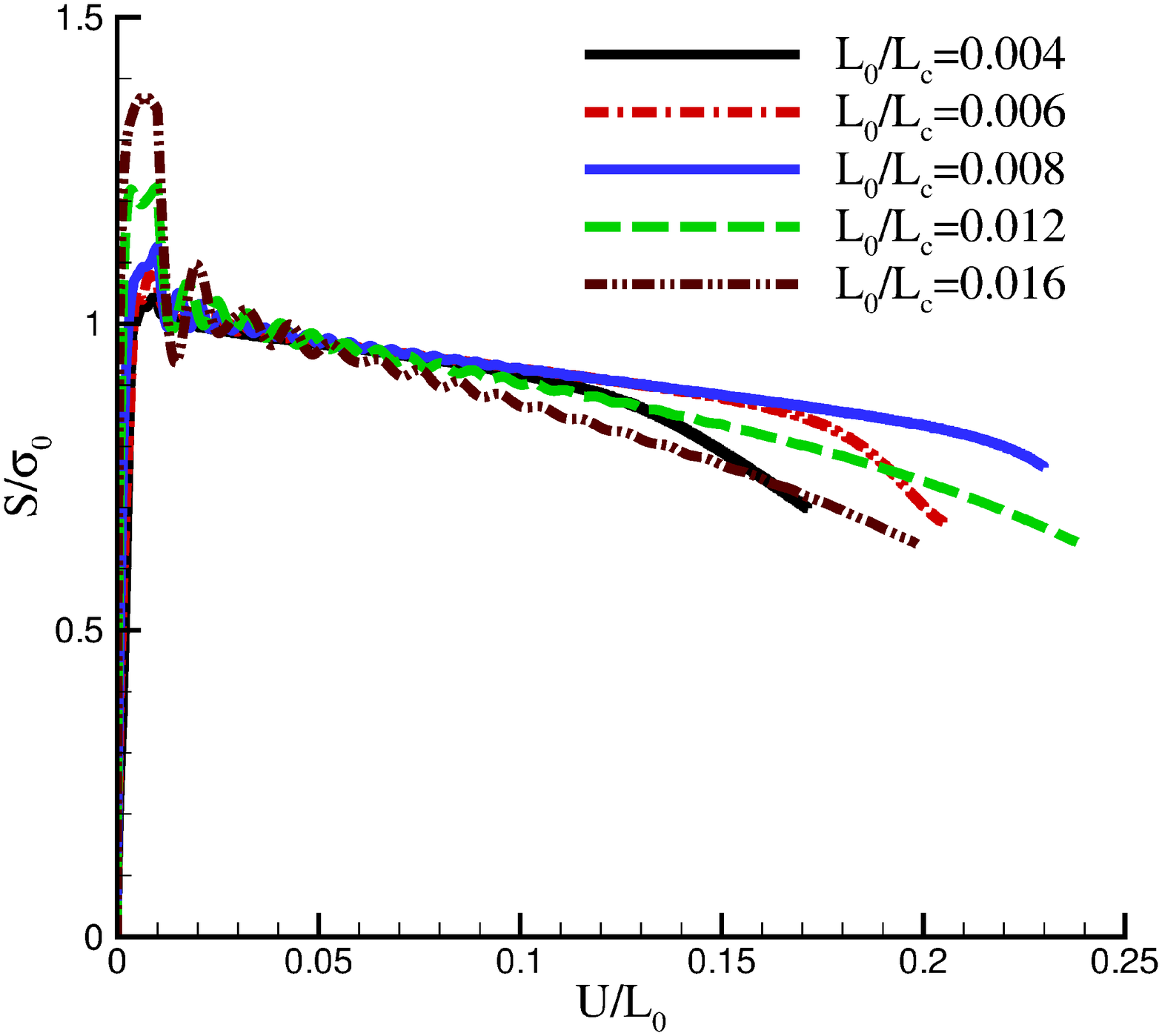}}}
\end{center}
\caption{Nominal stress-strain curves. (a) $m=0.01, N=0.01$. (b)
  $m=0.05, N=0.01$.}
\label{L-LY-ld}
\end{figure}

\begin{figure}[htb!]
\begin{center}
\subfigure[]
{\resizebox{!}{60mm}{\includegraphics{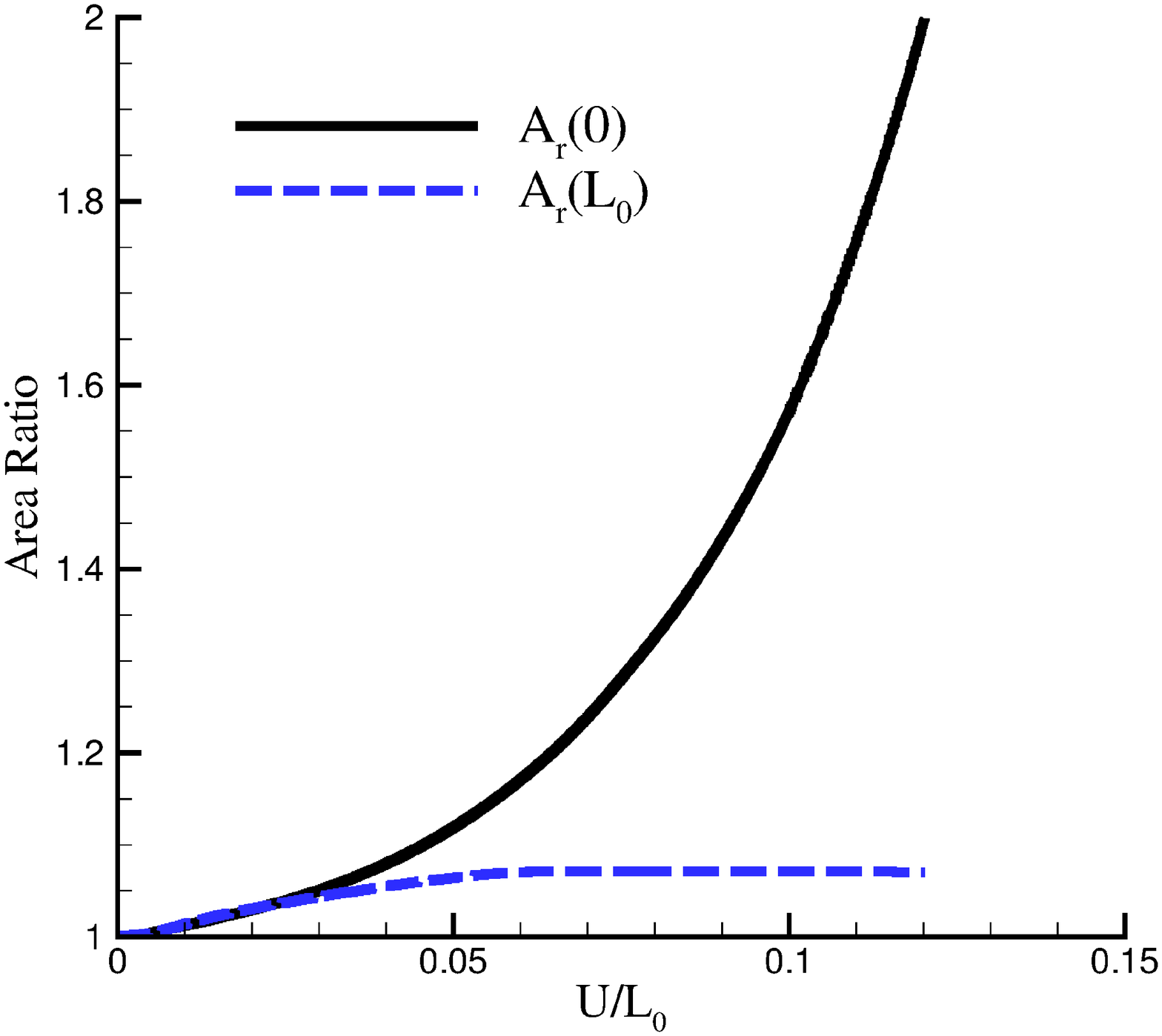}}}
\subfigure[]
{\resizebox{!}{60mm}{\includegraphics{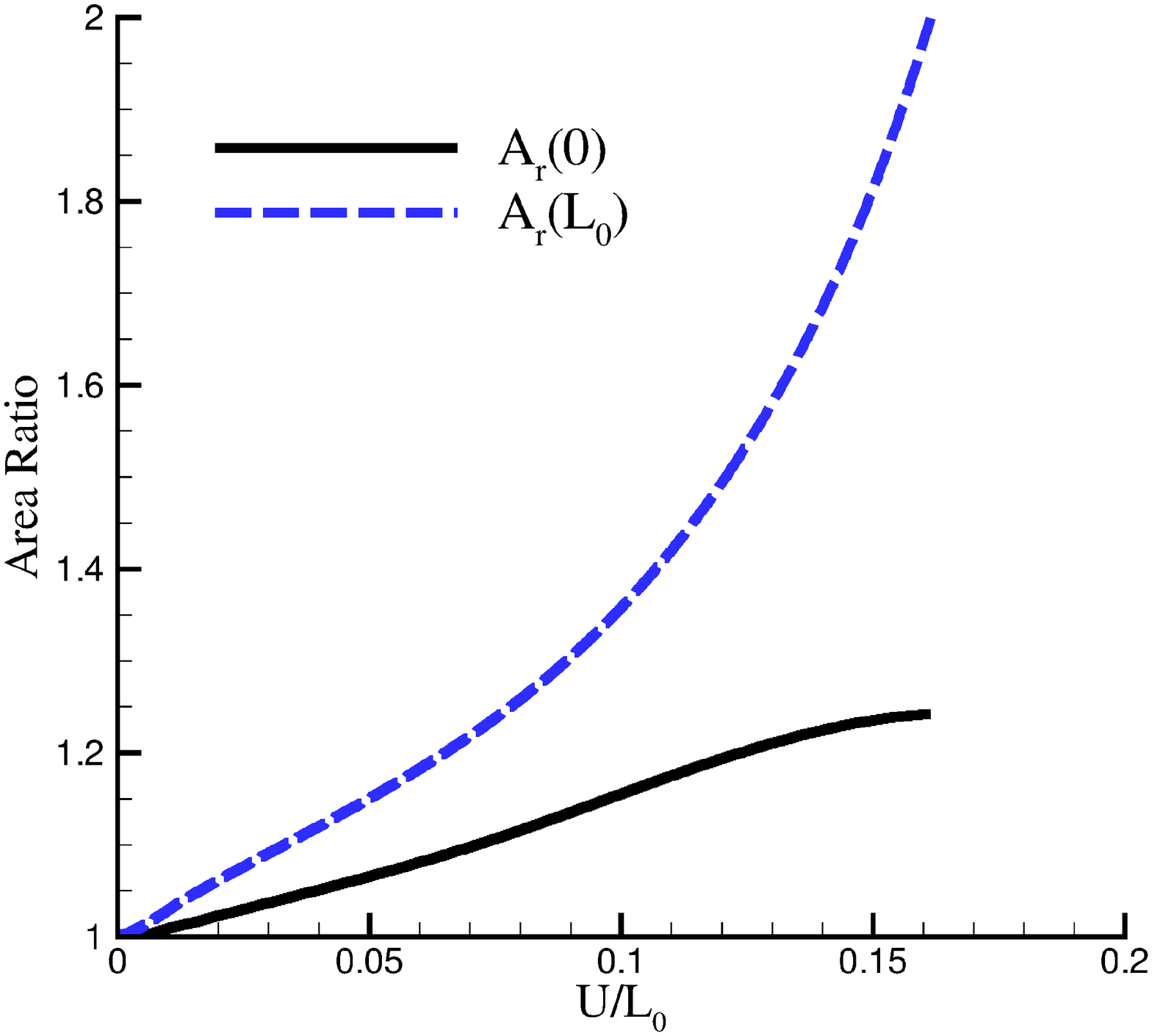}}}
\end{center}
\caption{Evolution of $A_r(0)$ and $A_r(L_0)$, with $A_r(z)$ given by
  Eq.~(\ref{area}). (a) $m=0.01, N=0.01$, $L_0/L_c=0.004$. (b)
  $m=0.01, N=0.01$, $L_0/L_c=0.012$.}
\label{ar2060}
\end{figure}

Fig.~\ref{L-LY-ld} shows nominal stress-strain curves for $m=0.01$,
Fig.~\ref{L-LY-ld}a, and for $m=0.05$, Fig.~\ref{L-LY-ld}b. In both
plots $N=0.01$. The quantity $S$ is the nominal stress, the force per unit reference
area, and the end displacement $U$ is $U=\int V dt$. For sufficiently
small $L_/L_c$ the response is essentially quasi-static and this is
nearly the case for $L_0/L_c=0.004$ in Fig.~\ref{L-LY-ld}. 

As $L_0/L_c$ increases from $0.004$, inertia plays an increasing role
which delays the onset of necking. The same 
overall trend occurs for both $m=0.01$ and $m=0.05$.  In
Fig.~\ref{L-LY-ld}a where $m=0.01$ the maximum 
necking strain occurs for $L_0/L_c=0.008$.  In Fig.~\ref{L-LY-ld}b
where $m=0.05$ the maximum necking strain occurs for
$L_0/L_c=0.012$. Thus, increased strain rate sensitivity results in
the maximum necking strain (as defined here) occurring for a  larger
value of $L_0/L_c$.  For the larger values of
$L_0/L_c$ wave effects come into play which is what leads to the
reduction in necking strain as also seen by \cite{KN93}. 

Fig.~\ref{ar2060} shows the evolution of the area ratios $A_r(0)$ and
$A_r(L_0)$ with strain, $U/L_0$, for two cases. (Recall that $z$
denotes a convected 
coordinate so that $z=L_0$ corresponds to the impact end in the
current configuration.) In Fig.~\ref{ar2060}a, $L_0/L_c=0.004$ and
necking occurs at 
the notch cross section while in Fig.~\ref{ar2060}b, $L_0/L_c=0.012$
necking occurs at the impact end. In both cases, the area reduction
(recall that $A_r$ in Eq.~(\ref{area}) increases with increasing area
reduction) eventually increases rapidly at one end and remains nearly
constant at 
the opposite end.  In Fig.~\ref{ar2060}a the response is
quasi-static like and the relative area changes at $z=0$ and $z=L_0$ are
nearly the same initially. On the other hand in
Fig.~\ref{ar2060}b, where material inertia plays an important role,
significant deformation occurs at $z=L_0$ before much plastic
deformation takes place at $z=0$.
It can also be seen in Fig.~\ref{ar2060} from the
rapid increase in $A_r$ when necking occurs, that although the precise
value of the necking strain depends on the cut-off value chosen
($A_r=2$ here), the trends will not be sensitive to this value. 

\begin{figure}[htb!]
\begin{center}
\resizebox*{75mm}{!}{\includegraphics{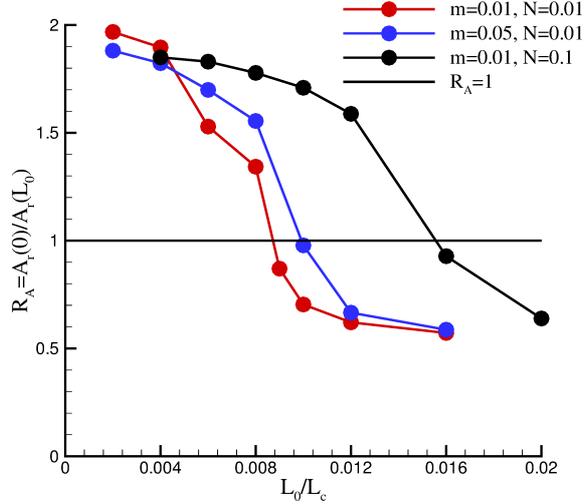}}
\end{center}
\caption{Ratio of the relative area reductions at $z=0$ and
  $z=L_0$, $R_A=A_r(0)/A_r(L_0)$. Values greater than $1$ correspond
  to necking at $z=0$ while 
  values less than $1$ correspond to necking at $z=L_0$.}
\label{arat}
\end{figure}

The values of $R_A$, defined in Eq.~(\ref{RA}), for all values of
$L_0/L_c$ considered
is shown in Fig.~\ref{arat}. The reference line $R_A=1$ corresponds to
simultaneous necking at the two ends of the region analyzed. For
$m=0.01$ necking occurs at the notch for $L_0/L_c$ less than about
$0.0084$ and occurs at the impact end for greater values of
$L_0/L_c$. For $m=0.05$ this transition occurs just about at
$L_0/L_c=0.01$. Increasing the strain hardening exponent from $N=0.01$
to $N=0.1$ leads to the transition
from necking at the notch to necking away from the notch taking place
at a much larger value of $L_0/L_c$. However, although strain hardening
and/or strain rate hardening can strongly affect the size, i.e. the
value of $L_0/L_c$, at which this transition in necking location
occurs, the results in Fig.~\ref{arat} show that it is inertia that drives
the transition. With $L_c=5$m as for the dimensional values given in Section
\ref{result}, the transition value of $L_0$ for $m=0.01$, $N=0.01$ is
$0.044$m while with $m=0.05$ this increases to $\approx 0.05$m and for
$m=0.01$, $N=0.1$, the transition to necking away from the notch takes
place for $L_0 \ge 0.078$m. 

\begin{figure}[htb!]
\begin{center}
\subfigure[]
{\resizebox{!}{60mm}{\includegraphics{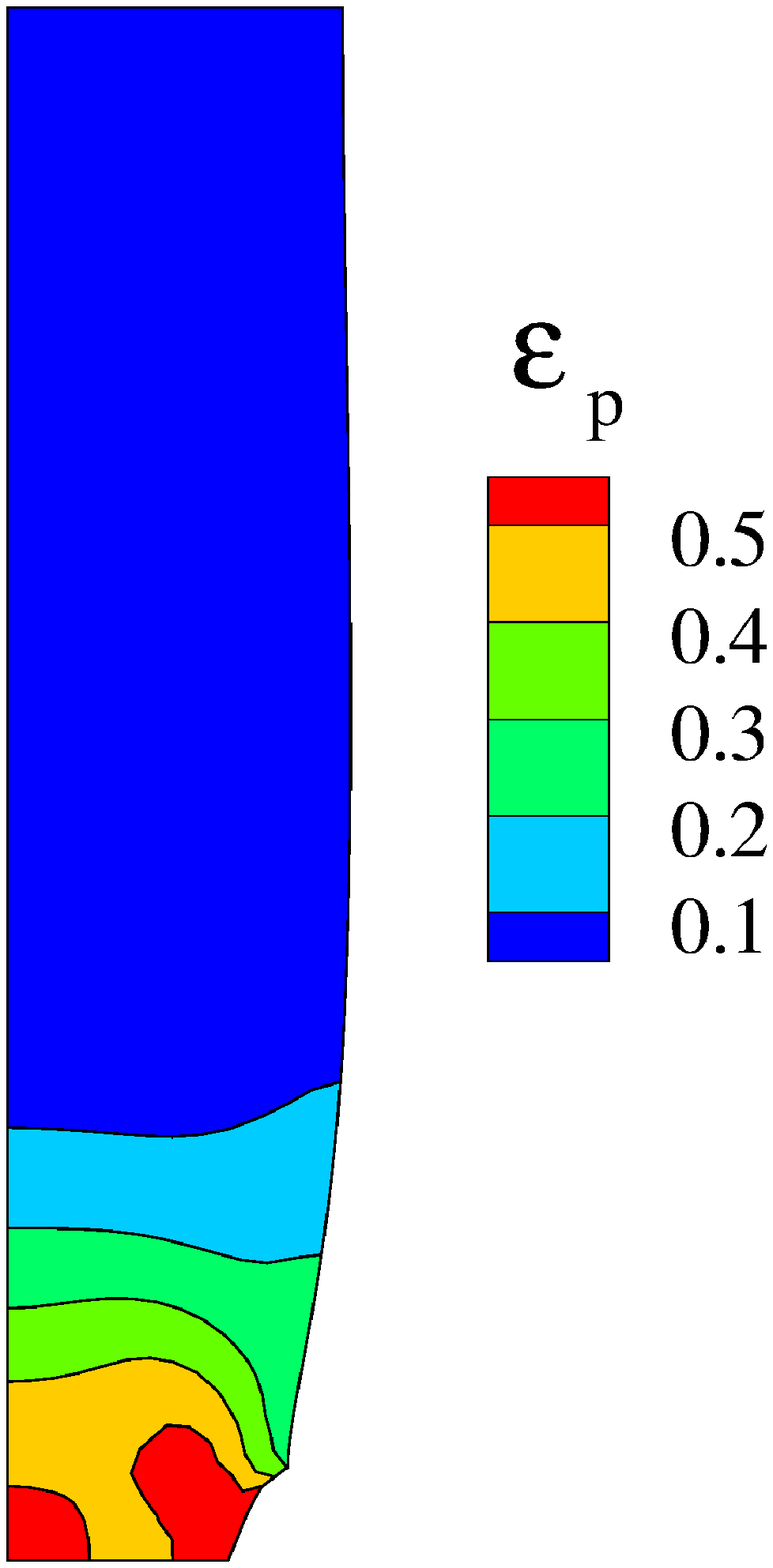}}}
\hspace{2cm}
\subfigure[]
{\resizebox{!}{60mm}{\includegraphics{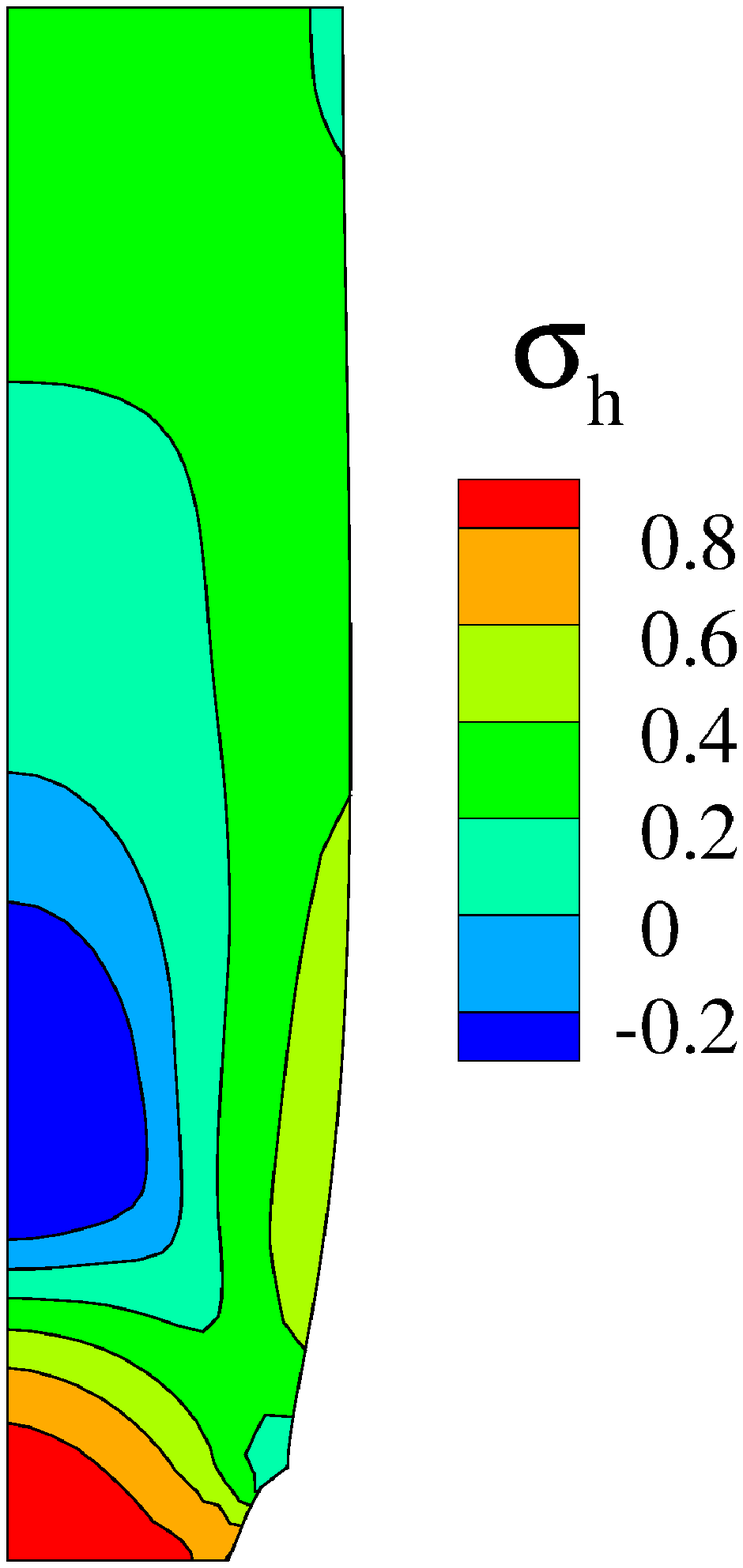}}}
\end{center}
\caption{(a) Distribution of effective plastic strain
  $\epsilon_p$. (b) Distribution of hydrostatic stress $\sigma_h$. For
$m=0.01$, $N=0.01$ and $L_0/L_c=0.004$ at $U/L_0=0.120$.}
\label{L20}
\end{figure}

Fig.~\ref{L20} shows contours of effective plastic strain, $\epsilon_p$, and
hydrostatic stress, $\sigma_h=\Sigma_h/\sigma_0$,  for a case where
necking occurs at the notch. There 
are two concentrations of plastic strain, $\epsilon_p$, in
Fig.~\ref{L20}a: one at the neck 
center and one that extends at about $45^\circ$ from the notch
root. The strain concentration that extends at the notch
root does not occur in a naturally necked specimen. The distribution
of  $\sigma_h$ has a hydrostatic
tension peak of about $0.9$ at the neck center and
hydrostatic compression away from 
the neck along the bar axis that reaches $-0.6$. The
distributions of $\epsilon_p$ and $\sigma_h$ are very similar to
those that would be obtained from a quasi-static
analysis. 

\begin{figure}[htb!]
\begin{center}
\subfigure[]
{\resizebox{!}{60mm}{\includegraphics{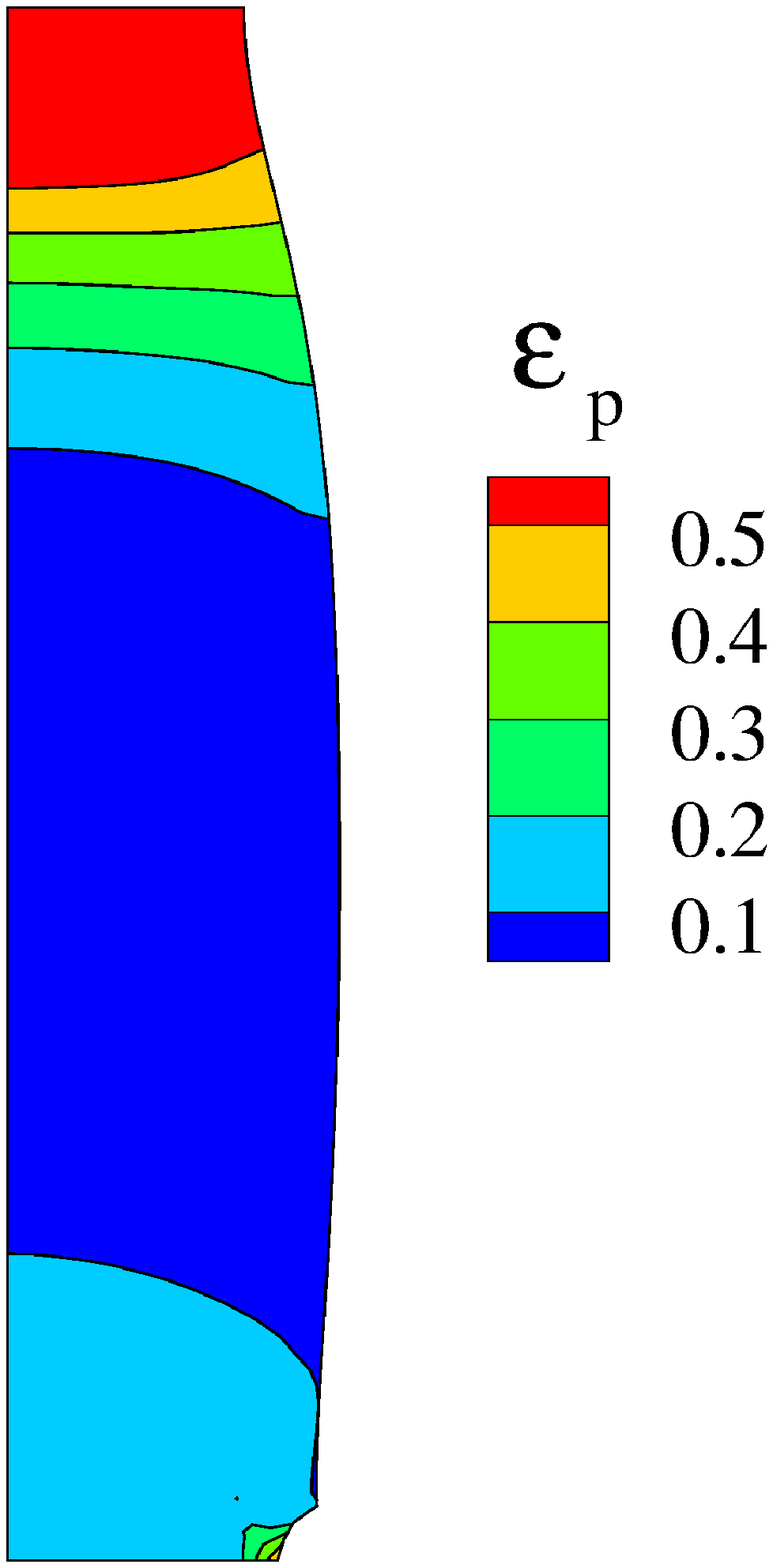}}}
\hspace{2cm}
\subfigure[]
{\resizebox{!}{60mm}{\includegraphics{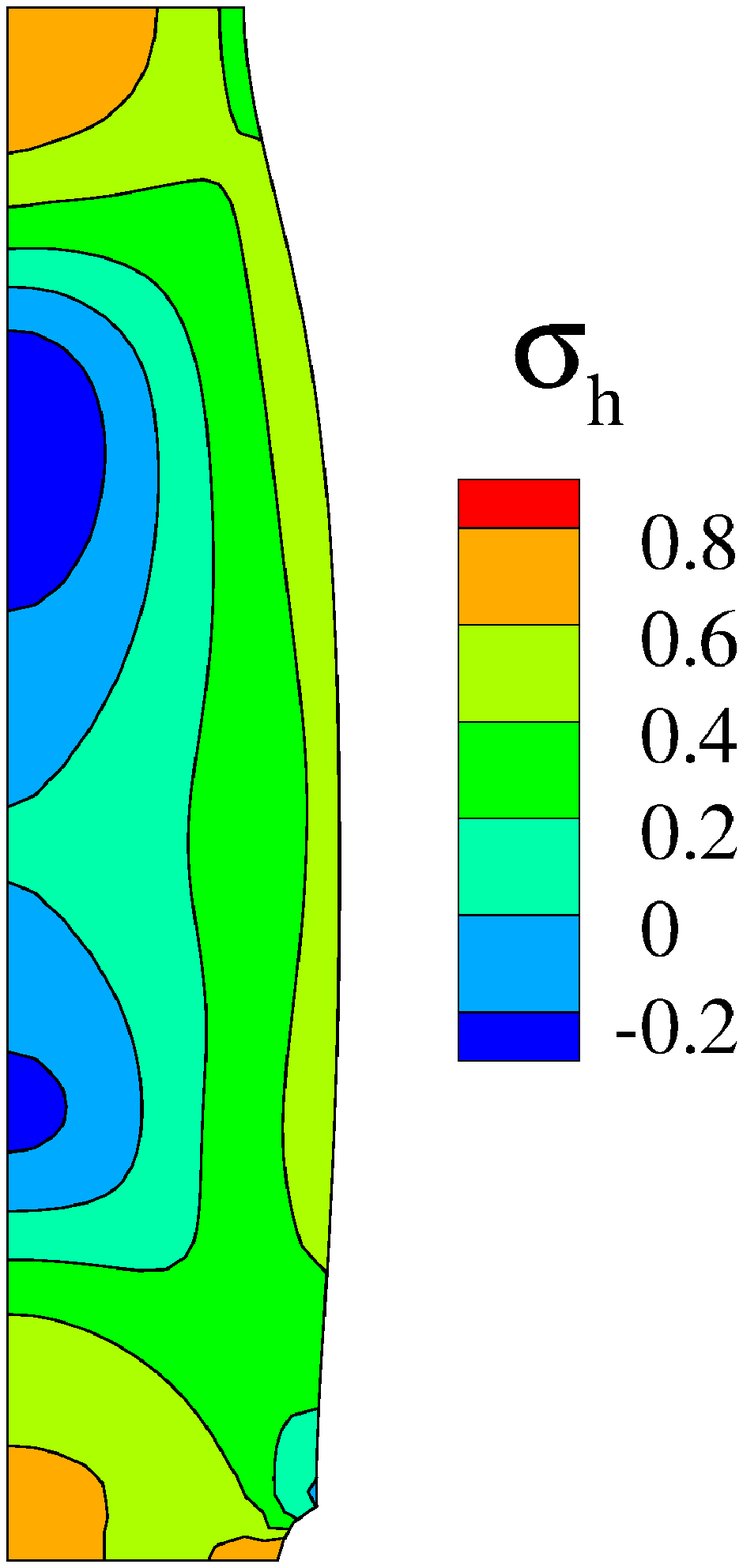}}}
\end{center}
\caption{(a) Distribution of effective plastic strain
  $\epsilon_p$. (b) Distribution of hydrostatic stress $\sigma_h$. For
  $m=0.01$, $N=0.01$ and $L_0/L_c=0.012$ at $U/L_0=0.161$.}
\label{L60}
\end{figure}

With $L_0/L_c=0.012$ in Fig.~\ref{L60} necking has occurred at the
impact end. As can be 
seen in Fig.~\ref{L60}a a strain concentration did initiate at the 
notch root and reach $\epsilon_p \approx 0.3$, but the deformation
eventually concentrated at the impact end. The distribution of
$\sigma_h$ in 
Fig.~\ref{L60}b shows evidence of the initial neck development at
$z=0$. There are three concentrations of hydrostatic tension; at the
notch root, at the center of
the bar at $z=0$ and at the center of the bar at $z=L_0$. The maximum
positive hydrostatic stress is in the
neck that forms at $z=L_0$, reaching $0.79$. There is also
a region of 
negative $\sigma_h$ along the axis as a consequence of the initial
neck formation there. The distributions of plastic strain $\epsilon_p$
and $\sigma_h$ in Fig.~\ref{L60} are very different from what would be
obtained from a quasi-static analysis.

\begin{figure}[htb!]
\begin{center}
\subfigure[]
{\resizebox{!}{60mm}{\includegraphics{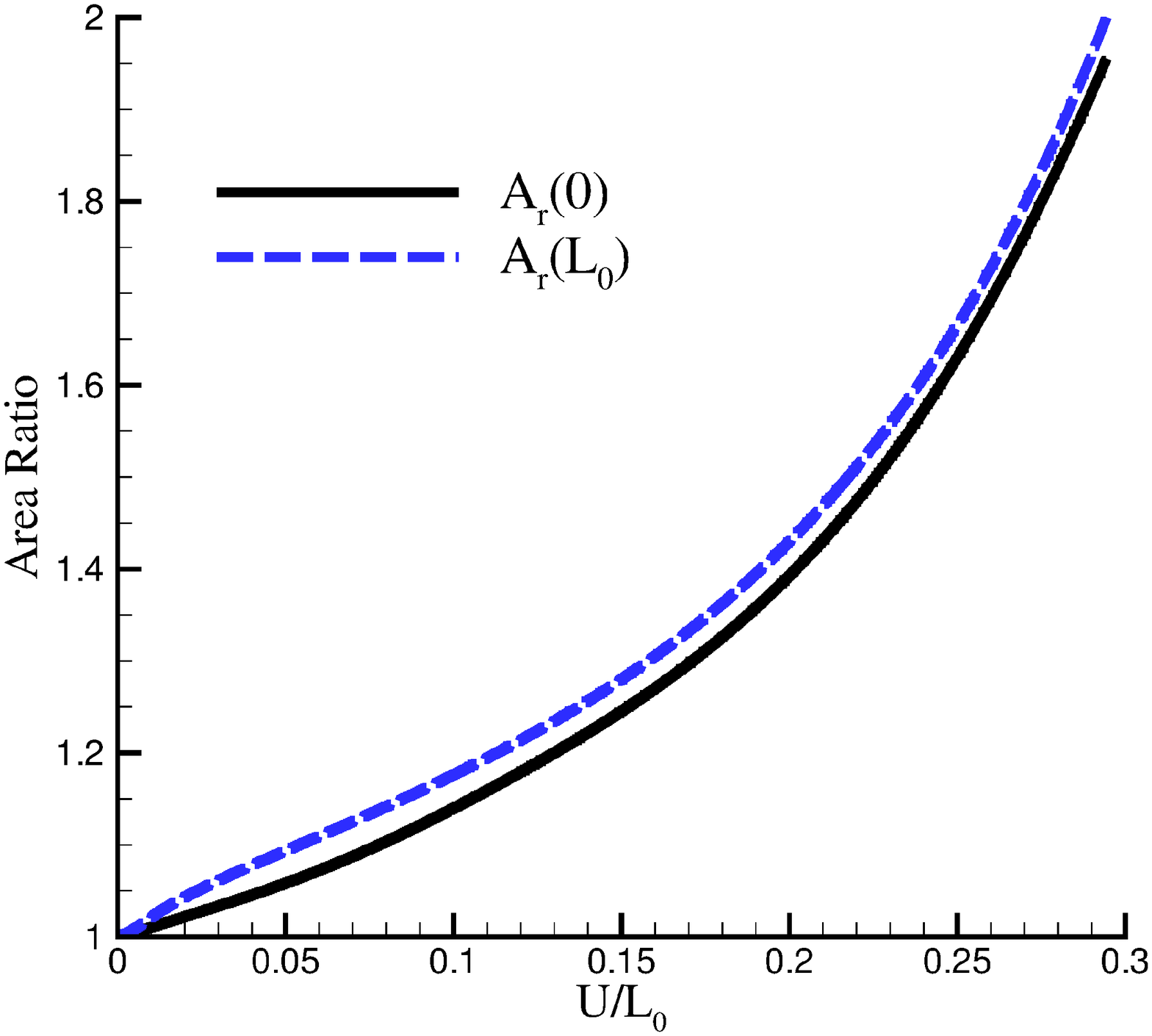}}}\\
\subfigure[]
{\resizebox{!}{60mm}{\includegraphics{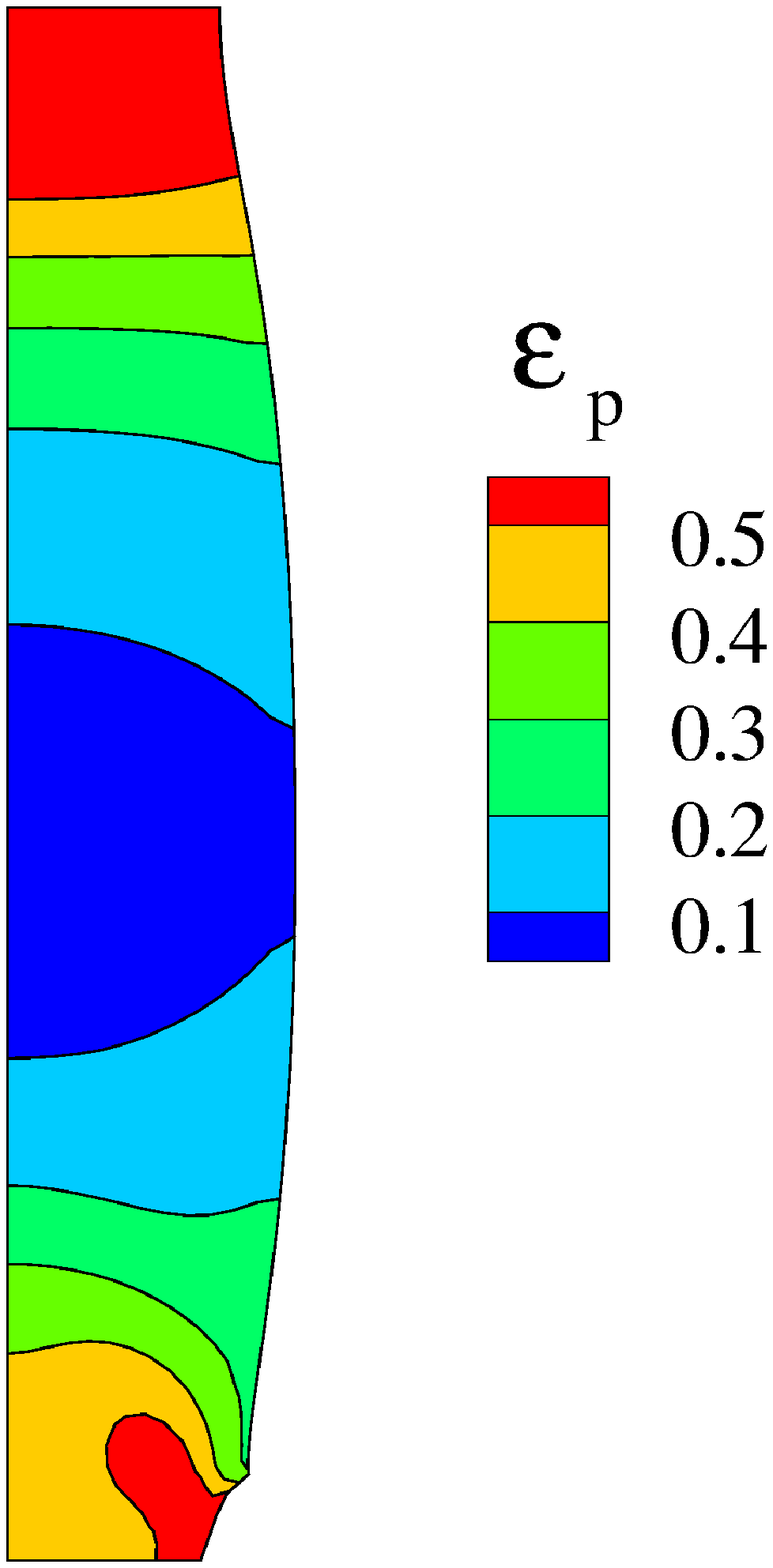}}}
\hspace{2cm}
\subfigure[]
{\resizebox{!}{60mm}{\includegraphics{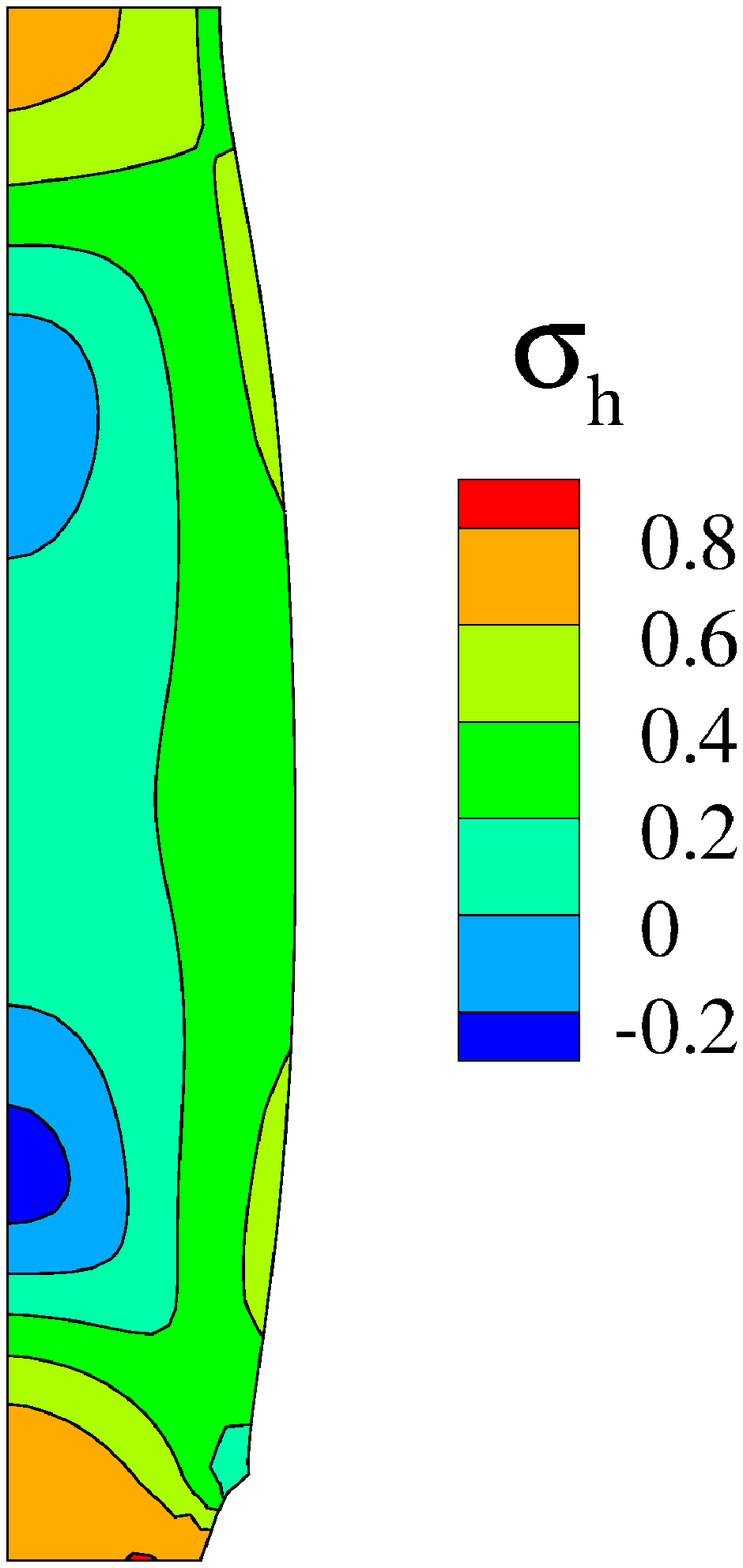}}}
\end{center}
\caption{(a) Evolution of the area ratios at $z=0$ and $z=L_0$, $A_r(0)/A_r(L_0)$. (b)
  Distribution of effective plastic strain 
  $\epsilon_p$. (c) Distribution of hydrostatic stress $\sigma_h$. For
$L_0/L_c=0.010$, $m=0.05$, $N=0.01$ at $U/L_0=0.294$.}
\label{LY50}
\end{figure}

The calculation with $m=0.05$, $N=0.01$ and $L_0/L_c=0.010$ is unusual
in that necking 
occurred nearly simultaneously at the notch plane ($z=0$) and at the
loaded end ($z=L_0/2$). Fig.~\ref{LY50}a shows the evolution of the
area ratios $A_r(0)$ 
and $A_r(L_0)$ with strain, $U/L_0$. Although
$A_r(L_0)$ grows slightly faster from the beginning, the two grow at
nearly the same rate and when  $A_r(L_0)=2$, the area ratio at the
$z=0$ is $A_r(0)=1.96$. The strain
distribution in the vicinity of the notch in Fig.~\ref{LY50}b is very
similar to that in Fig.~\ref{L20} with the strain concentration from
the notch root emanating at about $45^\circ$. In Fig.~\ref{LY50}b the plastic
strain near the notch root exceeds $1.0$ but this strain concentration
is very localized. The value of $\epsilon_p$ in the neck at $z=L_0$ is
about $0.7$ over a fairly large region. The hydrostatic stress
distribution in Fig.~\ref{LY50}c is nearly symmetrical consistent with
the nearly equal necking at $z=0$ and $z=L_0$.

To give an indication of the length scales involved, with, from
Eq.~(\ref{eq:L}), $L_c=c_0/(V_1/L_0)=5$m 
$L_0/L_c=0.01$ corresponds to $5$cm so that the transition from
necking at $z=0$ to $z=L_0$ occurs for $L_0$ between about $4$cm and
$8$cm.

\subsection{Fixed $L_0/L_c$, varying $\kappa$}

\begin{figure}[htb!]
\begin{center}
\subfigure[]
{\resizebox{!}{75mm}{\includegraphics{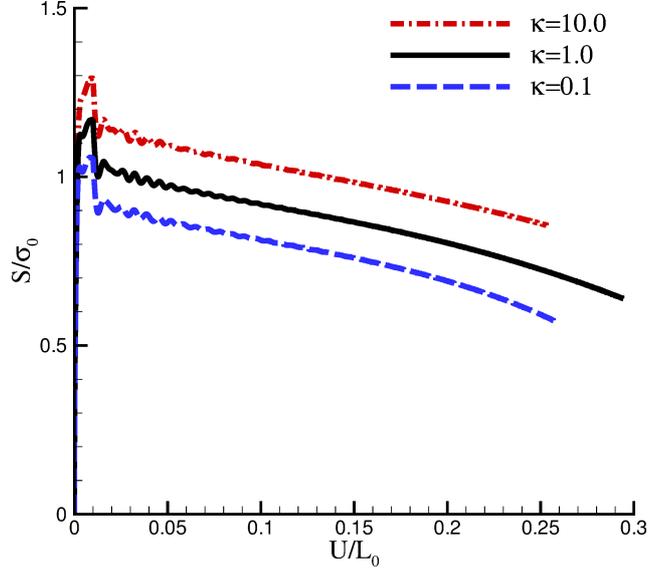}}}
\hspace{2cm}
\subfigure[]
{\resizebox{!}{75mm}{\includegraphics{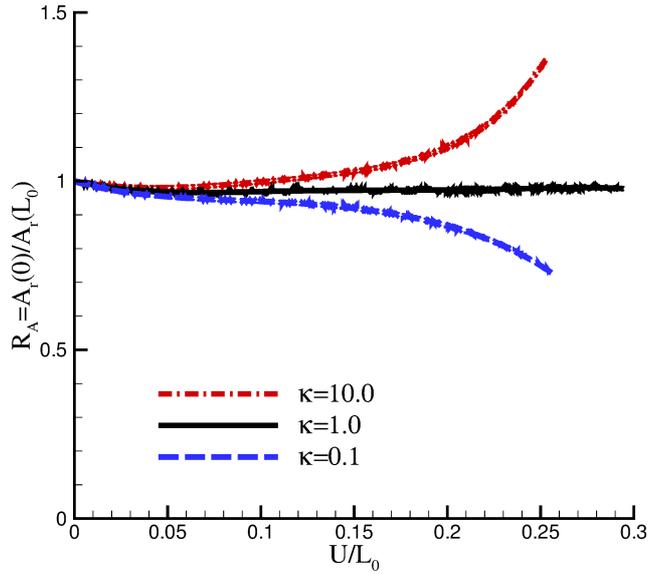}}}
\end{center}
\caption{(a) Stress strain curves.  (b) Evolution of $R_A$ defined in
  Eq.~(\ref{RA}). For 
$L_0/L_c=0.010$, $m=0.05$ and various values of $\kappa$.}
\label{LY50-ld}
\end{figure}

\begin{figure}[htb!]
\begin{center}
\subfigure[]
{\resizebox{!}{60mm}{\includegraphics{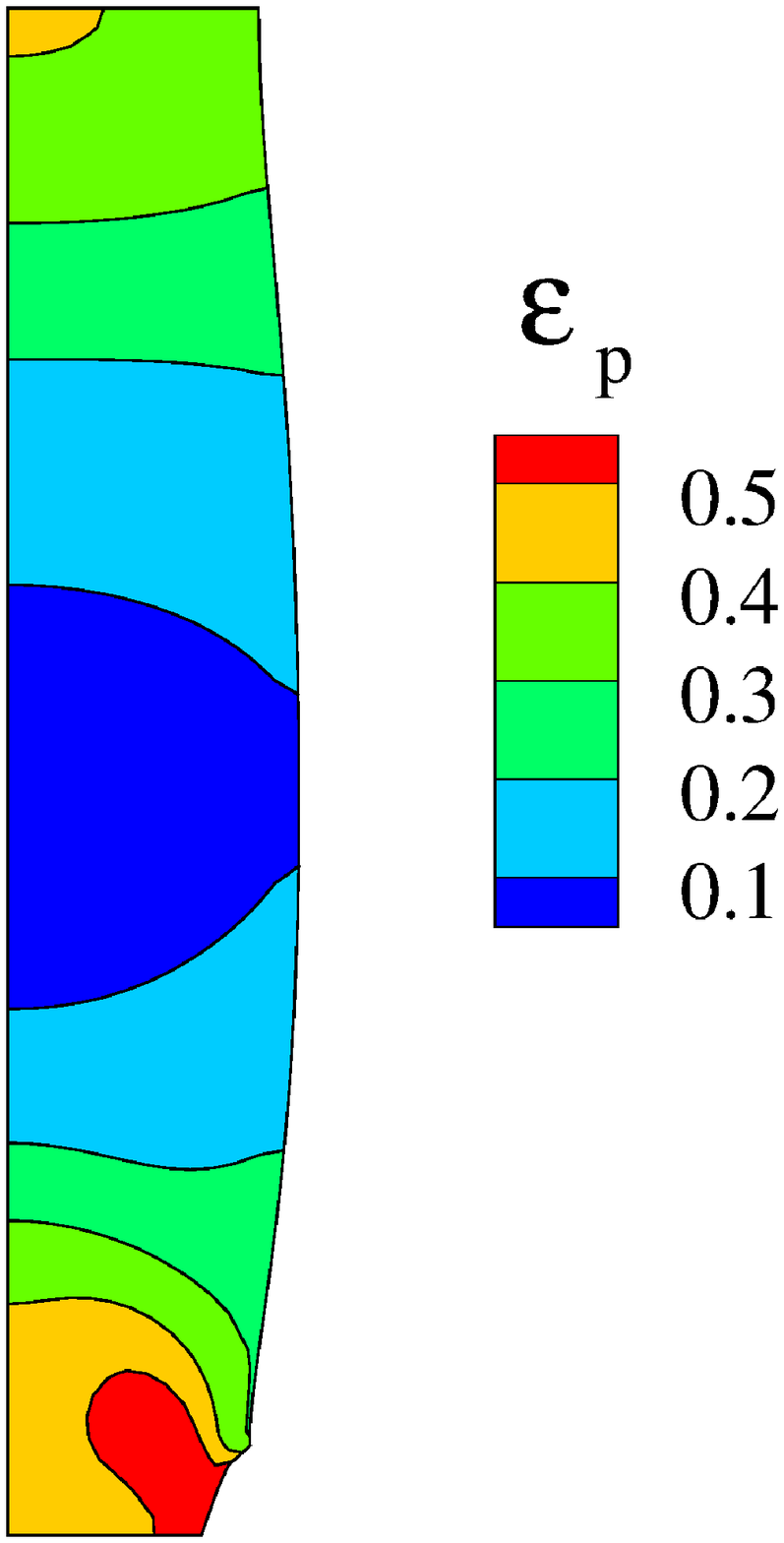}}}
\hspace{2cm}
\subfigure[]
{\resizebox{!}{60mm}{\includegraphics{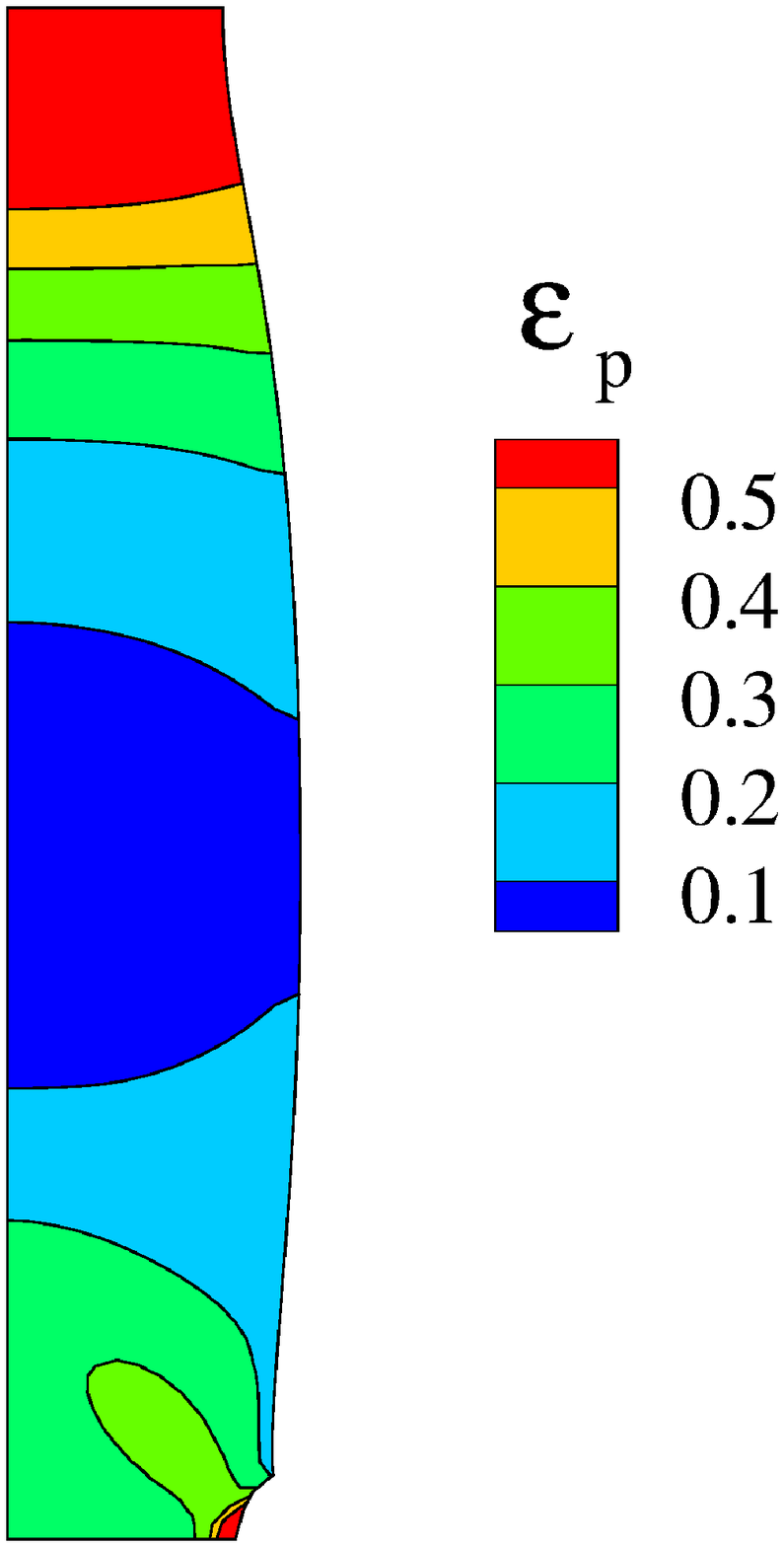}}}
\end{center}
\caption{(a) Distribution of effective plastic strain
  $\epsilon_p$ for $L_0/L_c=0.010$, $m=0.05$. (a) $\kappa=10$ at
  $U/L_0=0.253$. (b) $\kappa=0.1$ at $U/L_0=0.257$.} 
\label{LY50-ep}
\end{figure}

The calculations in this section are carried out for $L_0/L_c=0.01$ and
$m=0.05$. In Fig.~\ref{arat}, where $\kappa=1$, this is the case for
which necking occurs nearly simultaneously at $z=0$ and
$z=L_0$. Fig.~\ref{LY50-ld} shows nominal stress-strain curves and the
evolution of $R_A$ with imposed strain, $U/L_0$, for $\kappa=10$, $1$
and $0.1$. Larger values of $\kappa$, see Eq.~(\ref{eq:kap}),
correspond to larger imposed strain rates 
relative to the material strain rate $\dot{\epsilon}_0$. 

From Eq.~(\ref{eq:ratx}), $t_c \dot{\epsilon}_0 \rightarrow 0$ as
$\kappa \rightarrow \infty$ and $t_c \dot{\epsilon}_0 \rightarrow
\infty$ as $\kappa \rightarrow 0$. With $t_c \dot{\epsilon}_0 $ the ratio of
inertial and material time scales, this indicates that inertia dominates as
$\kappa \rightarrow 0$ and more quasi-static type behavior occurs as 
$\kappa \rightarrow \infty$. With $\dot{\epsilon}_0$ non-zero and finite,
this means that increased the imposed strain rate can correspond to more
quasi-static like behavior, which may seem counter intuitive. 

To understand this consider $\dot{\epsilon}_0$ fixed and $V_1/L_0$
varying. Increasing $\kappa$, Eq.~(\ref{eq:kap}), then implies
increasing  $V_1/L_0$, but 
since $L_0/L_c$ fixed, $V_1/c_0$ is fixed. For $V_1/L_0$ to increase
with $V_1/c_0$ unchanged requires $L_0$ to decrease. Since 
with $L_0/L_c$ fixed, the stress carried by the loading
wave is fixed, Eq.~(\ref{rel4}), this scenario
corresponds to a smaller specimen with a fixed loading wave
stress. Note also that varying $V_1/L_0$ with $\dot{\epsilon}_0$
fixed, implies that $t_c$ varies, see Eq.~(\ref{rel3}), so that to
obtain the properly 
scaled initial/boundary value problem, $T_r$ must be varied so that
$t_r=T_r/t_c$ remains fixed.

Fig.~\ref{LY50-ld}a shows stress-strain plots for $\kappa=10$, $1$ and
$0.1$. The calculation for $\kappa=1$ is the one discussed in Section
\ref{Lc} where necking occurred nearly simultaneously at $z=0$ and
$z=L_0$. The stress strain curves for all three values of $\kappa$ are
qualitatively similar with the stress levels varying as expected due
to the material strain rate sensitivity. It is worth noting that the
maximum strain to necking (i.e. to $A_r=2$ at some cross section)
occurs for the intermediate value $\kappa=1$.

Curves of the ratio $A_r(0)/A_r(L_0)$ are shown in
Fig.~\ref{LY50-ld}b. For $\kappa=1$, $A_r(0)/A_r(L_0) \approx 1$
consistent with the nearly simultaneous necking at $z=0$ and
$z=L_0$. For the increased relative strain calculation, $\kappa=10$,
$A_r(0)/A_r(L_0)>1$, indicating necking at $z=0$, while for the lower
strain rate, $\kappa=0.10$, $A_r(0)/A_r(L_0)<1$ indicating necking at
$z=L_0$. 

Contours of effective plastic strain, $\epsilon_p$, for $\kappa=10$
and $\kappa=0.1$ are shown in Fig.~\ref{LY50-ep} showing neck
development at $z=0$ for the higher relative strain rate and at
$z=L_0$ for the lower relative strain rate.

The value of $\kappa$ can be varied by keeping all material,
geometrical and loading parameters fixed except for
$\dot{\epsilon}_0$. In this situation, the results here show that the
where necking occurs can be sensitive to the value used for
$\dot{\epsilon}_0$. Thus, the predicted response can depend
qualitatively, not only quantitatively, on the value of
$\dot{\epsilon}_0$. 

\section{Discussion}

With fixed material properties and bar geometry, the results show that
with a fixed imposed strain rate, $V_1/L_0$, a transition from notch triggered
quasi-static like necking to  notch ignoring dynamic necking  occurs with
increasing bar size, $L_0/L_c$. Thus with $\kappa$ fixed, the effect of
material inertia on necking location increases with increasing values
of the imposed velocity $V_1$.  On the other hand, with $L_0/L_c$
fixed (with fixed bar geometry and with fixed material properties) but
with $\kappa$ varying, a similar transition takes place for a decreasing value of
the imposed strain rate $V_1/L_0$.  In both cases, the effect of material inertia
increases with increasing bar size $L_0$.

The idea that there can be a critical imposed
velocity  under dynamic loading conditions beyond which the apparent 
ductility decreases dates back to the 1930s, see for example
\cite{Mann36,K42,KN93,Kl05,Vaz16}. This critical velocity is associated with
material softening and often attributed to adiabatic heating. In the
circumstances analyzed here there is a
critical value of $L_0/L_c$ beyond which the strain
for necking decreases as seen in Fig.~\ref{L-LY-ld}. Since,  for fixed
material properties, 
$L_0/L_c \propto V_1$, there is a critical imposed velocity beyond
which the necking strain decreases. This
non-monotonic behavior is a consequence of material inertia; there
is no material softening in the formulation.

As seen in Fig.~\ref{LY50-ld}b the qualitative nature of the response
can depend on value of the parameter
$\kappa$ that couples the loading rate and the constitutive response
through the material parameter $\dot{\epsilon}_0$,
see  Eq.~(\ref{con5}). Thus, it is worth noting that, for a given value of the
imposed strain rate, the predicted 
response can strongly depend on the specified value of
$\dot{\epsilon}_0$. 

The parameter
$\kappa$,  the ratio between time scales associated 
with the material and the loading, does not enter a rate independent
formulation but does enter the formulation
for any value of the rate hardening exponent $m \ne 0$. Presumably,
the rate independent limit is somehow approached as $m \rightarrow
0$; one possibility is that the applied strain at which the responses
separate in Fig.~\ref{LY50-ld}b approaches infinity as $m \rightarrow
0$. In any case, the dependence of the necking behavior on $\kappa$ as $m
\rightarrow 0$ remains to be investigated.  Note
that even in the quasi-static
limit,  $\kappa$ can play a significant role, as
seen by \cite{Eran16} in a context different from the one considered
here. 

The non-dimensional ratio $L_0/L_c$, Eq.~(\ref{rel2}), does enter in the rate
independent limit as does the characteristic time $t_c$,
Eq.~(\ref{rel3}), which then 
only serves as a quantity for normalizing the time,
$t=T/t_c$. 

In the calculations here only one notch geometry was considered, a
semi-circular notch of radius $R_0/10$. It is expected that the
transition from necking at the notch cross section to necking at
another cross section will depend on the depth of the notch and,
probably to a lesser extent, on the shape of the notch. The results in
\cite{KN93} show that such a transition  occurs (in some case with necking taking
place at an intermediate location) even for naturally
necking bars with very small
geometrical imperfections, but the variation of the transition bar size with
increasing notch depth remains to be investigated. Also, the material
model used in the present calculations excludes any softening
mechanism, such as thermal softening and porosity induced
softening, which were included in the calculations in \cite{KN93}. The
dependence of the necking location transition seen here on such 
softening mechanisms, as well as on other aspects of the constitutive
characterization of the material remains to be explored.

Fineberg and co-workers, e.g. \cite{jay15}, have used soft materials
with slow wave speeds to study dynamic fracture processes at low
velocities in order to observe aspects of the crack growth process
that would be difficult or impossible to observe directly in hard
materials. The scaling properties embodied in the non-dimensional
equations here suggest that it may be possible also to do this to
study dynamic plastic instabilities, provided of course, that both the
soft and hard materials can be characterized using the same
elastic-viscoplastic constitutive framework.

\section{Conclusions}

The governing equations for dynamic  deformations of an
elastic-viscoplastic 
notched bar subject to tensile loading were presented in
non-dimensional form. Axisymmetric calculations were carried out for geometrically
identical bars having various sizes. Two key non-dimensional groups were identified
that contain a parameter characterizing the bar size. The main focus
was on variations of
the values of these two parameters that can be regarded as corresponding to
variations in bar size, although other interpretations are possible
and were discussed. Attention was principally directed at the effect
of material inertia on whether necking developed at the notch cross
section or whether necking ultimately occurred away from the notch.

It was found that:

\begin{description}

\item{1.} With a fixed imposed strain rate, the applied strain to
  necking (as defined here) does not depend monotonically on size.

\item{2.} The transition from notch induced necking to notch ignoring
  necking depends on size and is driven by material inertia.

\item{3.} One of the key non-dimensional groups that
  involves a measure of bar size is the ratio of the imposed
  velocity to an elastic wave speed, the other is
  the ratio of the imposed strain rate to the material characteristic
  strain rate. The second non-dimensional group is absent for a rate
  independent solid. 

\item{4.} With a fixed imposed strain rate and a fixed elastic
  wave speed,  necking was notch induced for sufficiently small values
  of the imposed velocity (smaller bar sizes).

\item{5.} With a fixed imposed velocity and a fixed material characteristic
  strain rate, necking was notch induced for sufficiently large values
  of the imposed strain rate (smaller bar sizes).

\end{description}

Thus, smaller may be stronger, see e.g. \cite{fleck94,greer05}, but larger is more
dynamic.

\end{document}